\documentclass[secnumarabic,nobibnotes,aps,notitlepage]{revtex4-1}
\usepackage{graphicx}
\usepackage{floatrow}

\usepackage{hyperref}

\def\Title#1{\begin{center} {\Large #1 } \end{center}}
\def\Author#1{\begin{center}{ \sc #1} \end{center}}
\def\Address#1{\begin{center}{ \it #1} \end{center}}

\newenvironment{Abstract}{\begin{quotation}  }{\end{quotation}}
\newenvironment{Presented}{\begin{quotation} \begin{center} 
             PRESENTED AT\end{center}\bigskip 
      \begin{center}\begin{large}}{\end{large}\end{center} \end{quotation}}

\begin{document}
\vskip 1.5 in
\vfill
\Title{The paradox of coherent photoproduction in incoherent interactions}
\vfill
\Author{Spencer R. Klein}%
\Address{Nuclear Science Division, Lawrence Berkeley National Laboratory, 1 Cyclotron Road, Berkeley, CA 94720 USA}
\vfill
\begin{Abstract}

The Good-Walker (GW) paradigm relates coherent and incoherent exclusive reactions to the average target configuration and its nucleonic and partonic fluctuations. In it, coherent photoproduction occurs when the target remains in the ground state, while, in incoherent photoproduction, the target breaks up.  However, the GW paradigm fails to explain the observation of coherent vector meson photoproduction accompanied by nuclear breakup in ultra-peripheral collisions (UPCs) , and in peripheral relativistic heavy-ion collisions.  In the latter, hundreds of particles can be created. This writeup will explore this paradox, and also present an alternate, semi-classical approach toward coherent production: adding the amplitudes with an appropriate propagator.  The semi-classical approach explains the transverse momentum dependence of exclusive vector meson production in UPCs, but does not address the target final state.   I will address these two approaches, and suggest possible future work to resolve the paradox  \cite{Klein:2023zlf}.

\end{Abstract}

\begin{Presented}
DIS2023: XXX International Workshop on Deep-Inelastic Scattering and
Related Subjects, \\
Michigan State University, USA, 27-31 March 2023 \\
     \includegraphics[width=9cm]{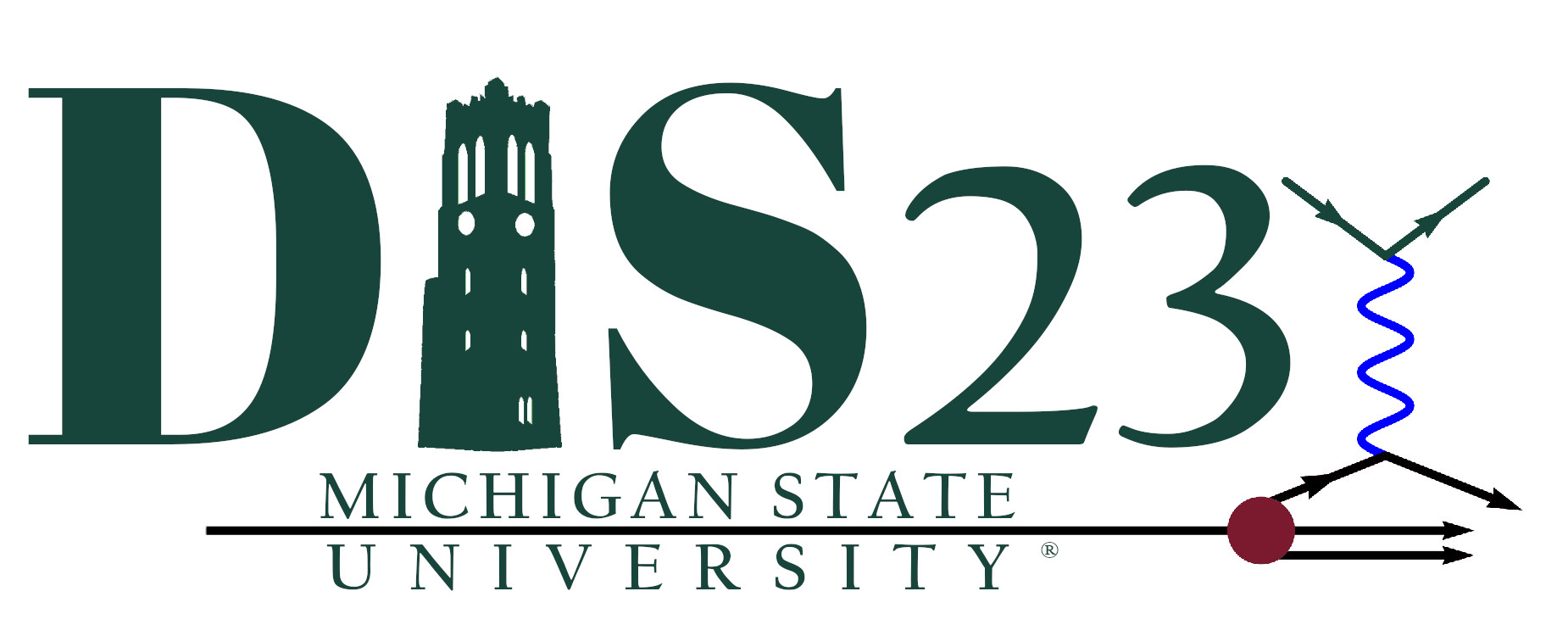}
\end{Presented}
\vfill
\eject 

{\bf Introduction}

Vector meson photoproduction is an important tool for studying nuclear structure, by probing gluon densities at low Bjorken$-x$. It is studied in ultra-peripheral collisions (UPCs) \cite{Bertulani:2005ru,Baltz:2007kq,Contreras:2015dqa}
and is a key part of the physics program at a future electron-ion collider (EIC) \cite{Accardi:2012qut,
AbdulKhalek:2021gbh}.  In coherent vector meson photoproduction, the $p_T$ distributions can be used to study the transverse distribution of nucleons in heavy nuclei \cite{STAR:2017enh,Klein:2019qfb}, targets, while measurements of incoherent photoproduction of these same mesons are sensitive to nucleonic and subnucleonic fluctuations.    An incident photon fluctuates to a long-lived $q\overline q$ dipole which then scatters elastically, via Pomeron exchange, from the target nucleus. 

Good and Walker \cite{Good:1960ba} and then Mietenlin and Pumplin \cite{Miettinen:1978jb,Frankfurt:2022jns} developed equations to relate coherent and incoherent diffraction respectively to the average nuclear configuration and its fluctuations.  The total cross-section for diffractive processes like photoproduction is \cite{Klein:2019qfb}
\begin{equation}
\frac{d\sigma_{\rm tot}}{dt} = \frac{1}{16\pi}\big<|A(K,\Omega)|^2\big>
\label{eq:tot}
\end{equation}
where $A(K,\Omega)$ is the amplitude for the process,  $K$ are the kinematic factors in the reaction (such as the photon and vector meson 4-momenta), and $\Omega$ is the instantaneous target configuration (such as the nucleon positions and partonic fluctuations).  The amplitudes are squared to get the cross-section for that configuration, and then averaged over configurations.   

In GW, in coherent interactions the target nucleus remains in its ground state. The  amplitudes are added, and
\begin{equation}
\frac{d\sigma_{\rm coh}}{dt} = \frac{1}{16\pi}\big|\langle A(K,\Omega)\rangle\big|^2.
\label{eq:coh}
\end{equation}
The target remains in its ground state, which leads to a coherent enhancement since one adds amplitudes before squaring.   A two-dimensional Fourier transform of $d\sigma_{\rm coh}/dt$ gives $F(b)$, the transverse distribution of individual scatterers in the target.  

The incoherent contribution is the remainder after the coherent cross-section is subtracted from the total:
\begin{equation}
\frac{d\sigma_{\rm inc}}{dt} = \frac{1}{16\pi}\bigg[\big<|A(K,\Omega)|^2\big>-\big|\big<A(K,\Omega)\big>\big|^2\bigg].
\label{eq:inc}
\end{equation}
This difference, between the sum of squares and the square of sums, is a measure of event-by-event
 fluctuations.  The momentum transfer $\sqrt{|t|}$ is still related to the length scale for fluctuations, but the relationship is less direct than  for coherent production.  In the limit of a totally absorptive black disk, the fluctuations disappear, and the incoherent cross-section is zero.   
\vskip .05 in
{\bf Coherent photoproduction with nuclear breakup}

A coherent enhancement in the cross-section at low $p_T$ ($p_T < \hbar/R_A$, where $R_A$ is the nuclear radius) has been observed in two types of reactions where the target does not remain intact.   The first is vector meson photoproduction in UPCs, accompanied by nuclear dissociation, which has been seen by STAR and ALICE.  The main STAR UPC trigger requires neutrons from nuclear breakup to be visible in both zero degree calorimeters (ZDCs).  With this trigger, STAR has observed coherent photoproduction of the $\rho^0$, direct $\pi^+\pi^-$ and a $\rho'$ decaying to four pions \cite{STAR:2002caw,STAR:2007elq,STAR:2017enh,STAR:2009giy}.   
ALICE does not use ZDCs in their trigger, but it does use them to select different impact parameter ranges \cite{ALICE:2020ugp}.

\begin{figure}
\floatbox[{\capbeside\thisfloatsetup{capbesideposition={right,top},capbesidewidth=6cm}}]{figure}[\FBwidth]
{\caption{Schematic diagram of vector meson photoproduction accompanied by mutual Coulomb excitation.  The vertical dashed lines show that the photons are independent, sharing only a common impact parameter. From. Ref.  \cite{Klein:2023zlf}.}\label{fig:multiphoton}}
{\includegraphics[width=4cm]{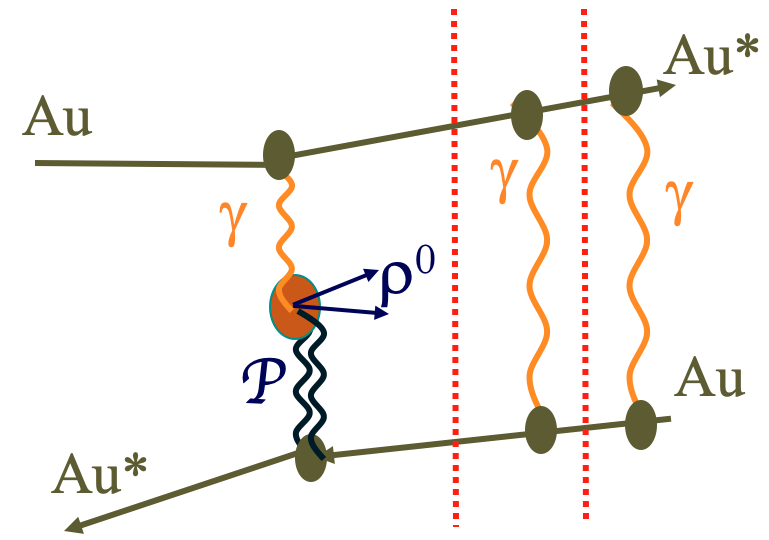}}
\end{figure}

These interactions can occur via multi-photon exchange (Fig. \ref{fig:multiphoton}).  Each photon does one thing, and the cross-section for a complex reaction is determined by integrating over $d^2b$ the product of the probabilities of each interaction occurring at that impact parameter.  ALICE has shown that this approach  predicts the abundances of different neutron classes (no neutrons, neutrons from one nucleus, or neutrons from both nuclei) accompanied by exclusive production of a $\rho^0$ at low $p_T$ \cite{ALICE:2020ugp}.  The GW approach, though, applies only to the total reaction, since it is insensitive to any internal structure.  

The second example is coherent photoproduction of the $J/\psi$ in peripheral relativistic heavy-ion collisions, leading to a large increase in $J/\psi$ production when $p_T < \hbar/R_A$ \cite{STAR:2019yox,ALICE:2022zso,LHCb:2021hoq}.  The hadronic parts of these collisions can create hundreds of particles.  There is no clear analog for the factorization like in Fig. \ref{fig:multiphoton}.  

Other subreactions are possible.  Although the amplitude is small, an ion can radiate a bremsstrahlung photon during photoproduction.  The photon energy ($k$) spectrum is $1/k$, so the cross-section is infrared divergent. With a low enough energy cutoff, radiation is present in every reaction.  Two-photon production of $e^+e^-$ pairs also has a large cross-section.  For a grazing lead-lead collision ($b\approx 2R_A$) at the LHC,  the average number of pairs is about 2.5, independent of what else happened during the encounter.  So, in many cases, vector meson production is accompanied by one or more $e^+e^-$ pairs.  Because the leptons typically have small $p_T$ and large rapidity, most of the pairs are invisible in current detectors.  However, their presence challenges GW.  

For Coulomb excitation, bremsstrahlung and pair production, it might possible to partially evade the problem by using an argument about their time scales.  These soft processes occur over time scales that are long compared to vector mesons production.  It may be argued that the vector meson production occurs before the nuclei dissociation occurs.  However, the GW paradigm does not consider time sequencing - only the particles before and after the complete interaction -  and, again, only strictly holds for stable particles \cite{Miettinen:1978xv}.  This argument does not apply for peripheral heavy-ion collisions.

\vskip .05 in
{\bf A semi-classical approach}

These data can be explained in a semiclassical approach.   As long as one cannot tell which nucleon was struck, the amplitudes for the photon/dipole to interact with the different nucleons in the target are added, so the cross-section is
\begin{equation}
    \sigma = \big|\Sigma_i^N A_i  e^{i\vec{k}\cdot\vec{x}_i}\big|^2.
    \label{eq:sigmasimp}
\end{equation}
The sum $i$ runs over the $N$ nucleons at positions $\vec{x}_i$.  The production amplitudes are $A_i$. and $k$ is the momentum transfer from the target to the vector meson.  We take the $A_i$ to be identical, but that is not strictly required.  

Equation \ref{eq:sigmasimp} naturally leads to a $p_T$ spectrum with two components:
\begin{equation}
\frac{d\sigma}{dt}= A_A \exp(-b_{A}t) + A_n \exp(-b_{n}t)
\label{eq:twocomponent}
\end{equation}
The cross-section $A_A$ and slope $b_A$ correspond to coherent interactions with the nucleus as a whole; the case where, in Eq. \ref{eq:sigmasimp}, $\vec{k}\cdot\vec{x}_i<1$, so the amplitudes add in phase, and $A_A \propto N^2$, and $b\approx (\hbar/R_A)^2$.  
The cross-section $A_n$ and slope $b_n$ correspond to incoherent photoproduction, where $\vec{k}\cdot\vec{x}_i \gg 1$, so the amplitudes add with random phase, so $A_n \propto N$.  The slope $b_n\propto (\hbar/R_p)^2$ corresponds roughly to the size of the individual nucleon targets. These regimes are observed in UPC data, with $p_T$ spectra in qualitative agreement with interactions from the entire nucleus and from individual nucleons \cite{STAR:2007elq}.
Equation \ref{eq:twocomponent} could be extended to include subnucleonic structure by adding a 3rd component, with a harder $t$ spectrum \cite{ALICE:2021gpt}.   For this, the sum in Eq. \ref{eq:sigmasimp} would run over the partons that can (energetically) contribute to the reaction. 
 
 It is also necessary to account for multiple interactions as a single photon/dipole passes through the target.
  This is usually done with a Glauber calculation \cite{Gribov:1968jf}. Multiple-interaction corrections reduce the cross-section, but do not alter the arguments presented here.  They reproduce the observed cross-sections for photoproduction of light vector mesons on heavy nuclei at low $p_T$ (in the coherent regime), either with or without excitation  \cite{Klein:1999qj,Frankfurt:2002sv}; the agreement improves by including excited intermediate states of the dipole \cite{Frankfurt:2015cwa}.  For the $J/\psi$ and $\psi'$, gluon shadowing must be included \cite{ALICE:2021gpt, Guzey:2013xba}.  
 
 For coherent UPC photoproduction, the semiclassical approach makes predictions that are similar to GW.  However, for incoherent production, the pictures are very different.    Semi-classically, there is no association between event-by-event fluctuations and the incoherent cross-section.  Instead, the incoherent cross-section and its $p_T$ spectrum depend on the number of emitters and their spatial arrangement.  A nucleus consisting of fixed, static nucleons would still interact incoherently.  In contrast, in GW,  without fluctuations, the incoherent cross-section is zero.  

 Equation \ref{eq:sigmasimp} can be applied to $J/\psi$ photoproduction in peripheral collisions, albeit with some open questions \cite{Zha:2017jch,Klusek-Gawenda:2015hja}: How many nucleons are involved in the coherence?  Is it all of the nucleons, or only the spectators?  Hadronic reactions occur on a similar or shorter time scale than photoproduction.  Does the targets interact hadronically, and lose energy before the photoproduction?  If so, the lower collision energy will reduce the photonuclear cross-section.   Both time orderings should be considered.  Unfortunately, the data is not yet precise enough to probe these questions.   
 
One assumption in the GW approach is that the incident projectile is a single photon (or other particle).   In UPCs, the strong electromagnetic fields permit multi-photon exchange, violating that assumption.  The presence of hadronically interacting particles in peripheral collisions similarly violates this assumption.  Multi-photon reactions such as in Fig. \ref{fig:multiphoton} are a subset of the one-photon reaction, and so absorb some of the one-photon cross-section, effectively changing coherent interactions into incoherent.   They also absorb some of the energy of the incident particle, contribute to the measured $t$, and so alter the apparent division between coherent and incoherent cross-sections.   In fact, it is impossible to tell if a reaction that produces a vector meson and excites a nucleus occurs via one-photon or two-photon exchange.  At  $p_T\approx 5 \hbar/R_A$ the amplitudes for one-photon and two-photon processes to produce a vector meson and excite the target are similar.    Alternately, the GW approach could be applied to the incident electron or ion, rather than a photon.   This would provide a single incident particle, but then it would clearly be impossible to treat the nuclear breakup as a separate subreaction.  

The semiclassical approach again finds support in the STAR $\rho$ photoproduction data, which measured the cross-section up to moderately large $|t|$.  In the range $0.2 < |t|<0.45$ GeV$^2$, the coherent cross-section is small, so only incoherent production should be present.  STAR found that $d\sigma/dt$ was well fit by a dipole form factor, which is commonly used to model a single proton  \cite{STAR:2017enh,Klein:2021mgd}, showing that, at least in this $|t|$ range,  the single-nucleon target is a good model.  The dipole form factor is also seen in color glass condensate calculations \cite{Mantysaari:2022sux}.   In GW, the incoherent cross-section is driven by fluctuations, and there is no reason it must follow a dipole form factor. 

\vskip .05 in
{\bf Other issues with $d\sigma_{\rm inc}/dt$ at low $|t|$}

There are other difficulties with $d\sigma_{\rm inc}/dt$ at low $|t|$,  some of which further challenge the GW approach.   One issue arises because nuclear excitation is endothermic, and as $|t|\rightarrow 0$ there is insufficient energy transferred to excite the nucleus, so incoherent interactions become impossible.   

For energy transfers above 5-8 MeV, heavy nuclei typically dissociate via proton or neutron emission \cite{Klein:2023zlf}.    If the Pomeron transfers its energy to a single nucleon, as indicated by the STAR data \cite{STAR:2017enh}, in order to transfer enough energy to cause nucleon emission, then the momentum transfer must be at least 100 to 125 MeV/c.   At substantially smaller Pomeron $p_T$, only excitation followed by photon emission is possible.  In this low-$p_T$ region $d\sigma_{\rm inc}/dt$ may be substantially smaller than an extrapolation from higher $|t|$ would indicate. 

Lower energy excitations must decay by photon emission.  In this regime, there are significant differences between the two commonly used nuclei, $^{208}$Pb and $^{197}$Au \cite{levels}. $^{208}$Pb is doubly magic, with a lowest lying excited state at 2.6 MeV, while $^{197}$Au has its lowest lying state at 77 keV.  This state has a 1.9 nsec lifetime, so, for a 110 GeV/n gold nucleus (planned for the EIC), the characteristic decay distance is about 70 m, long enough so that the decay occurs outside the detector, so is essentially invisible, and that event will be labelled as coherent. 

The different energy excitation spectra should lead to different cross-sections for incoherent vector meson production at low $|t|$; for some low-$|t|$ range, excitation of gold will be much easier than excitation of lead.  In contrast, in the GW paradigm, since they have similar sizes, density distributions (both are well-described by  Woods-Saxon distribution) and similar parton distributions and nuclear shadowing, they should have similar incoherent emission.  Comparing incoherent photoproduction with gold and lead targets at low $|t|$ will be an interesting test of the GW paradigm.  

\vskip .05 in
{\bf Conclusions}

The Good-Walker paradigm fails for two types of heavy-ion collisions:  coherent photoproduction accompanied by mutual Coulomb excitation in UPCs, and coherent photoproduction in peripheral heavy-ion collisions.   A semi-classical approach based on adding amplitudes can explain the coherent enhancement and $p_T$ distribution, but does not consider the final state of the nucleus.    These two approaches to coherent production paint profoundly different pictures of incoherent production.  Reconciling them is critical to understanding how incoherent production is sensitive to fluctuations in the average nuclear configuration.   A higher-order calculation might offer a way to reconcile this paradox.   

I  thank Guillermo Contreras Nuno, Daniel Tapia-Takaki, Mark Strikman, Heikki Mantysaari, Igor Pshenichnov, Lee Bernstein and the LBNL Nuclear Structure group for useful discussions.  This work is supported in part by the U.S. Department of Energy, Office of Science, Office of Nuclear Physics, under contract numbers DE-AC02-05CH11231.  

\bibliographystyle{apsrev4-1} 
\bibliography{main}

\begin{thebibliography}{30}%
\makeatletter
\providecommand \@ifxundefined [1]{%
 \@ifx{#1\undefined}
}%
\providecommand \@ifnum [1]{%
 \ifnum #1\expandafter \@firstoftwo
 \else \expandafter \@secondoftwo
 \fi
}%
\providecommand \@ifx [1]{%
 \ifx #1\expandafter \@firstoftwo
 \else \expandafter \@secondoftwo
 \fi
}%
\providecommand \natexlab [1]{#1}%
\providecommand \enquote  [1]{``#1''}%
\providecommand \bibnamefont  [1]{#1}%
\providecommand \bibfnamefont [1]{#1}%
\providecommand \citenamefont [1]{#1}%
\providecommand \href@noop [0]{\@secondoftwo}%
\providecommand \href [0]{\begingroup \@sanitize@url \@href}%
\providecommand \@href[1]{\@@startlink{#1}\@@href}%
\providecommand \@@href[1]{\endgroup#1\@@endlink}%
\providecommand \@sanitize@url [0]{\catcode `\\12\catcode `\$12\catcode
  `\&12\catcode `\#12\catcode `\^12\catcode `\_12\catcode `\%12\relax}%
\providecommand \@@startlink[1]{}%
\providecommand \@@endlink[0]{}%
\providecommand \url  [0]{\begingroup\@sanitize@url \@url }%
\providecommand \@url [1]{\endgroup\@href {#1}{\urlprefix }}%
\providecommand \urlprefix  [0]{URL }%
\providecommand \Eprint [0]{\href }%
\providecommand \doibase [0]{http://dx.doi.org/}%
\providecommand \selectlanguage [0]{\@gobble}%
\providecommand \bibinfo  [0]{\@secondoftwo}%
\providecommand \bibfield  [0]{\@secondoftwo}%
\providecommand \translation [1]{[#1]}%
\providecommand \BibitemOpen [0]{}%
\providecommand \bibitemStop [0]{}%
\providecommand \bibitemNoStop [0]{.\EOS\space}%
\providecommand \EOS [0]{\spacefactor3000\relax}%
\providecommand \BibitemShut  [1]{\csname bibitem#1\endcsname}%
\let\auto@bib@innerbib\@empty
\bibitem [{\citenamefont {Klein}(2023)}]{Klein:2023zlf}%
  \BibitemOpen
  \bibfield  {author} {\bibinfo {author} {\bibfnamefont {S.~R.}\ \bibnamefont
  {Klein}},\ }\href@noop {} {\  (\bibinfo {year} {2023})},\ \Eprint
  {http://arxiv.org/abs/2301.01408} {arXiv:2301.01408 [hep-ph]} \BibitemShut
  {NoStop}%
\bibitem [{\citenamefont {Bertulani}\ \emph {et~al.}(2005)\citenamefont
  {Bertulani}, \citenamefont {Klein},\ and\ \citenamefont
  {Nystrand}}]{Bertulani:2005ru}%
  \BibitemOpen
  \bibfield  {author} {\bibinfo {author} {\bibfnamefont {C.~A.}\ \bibnamefont
  {Bertulani}}, \bibinfo {author} {\bibfnamefont {S.~R.}\ \bibnamefont
  {Klein}}, \ and\ \bibinfo {author} {\bibfnamefont {J.}~\bibnamefont
  {Nystrand}},\ }\href {\doibase 10.1146/annurev.nucl.55.090704.151526}
  {\bibfield  {journal} {\bibinfo  {journal} {Ann. Rev. Nucl. Part. Sci.}\
  }\textbf {\bibinfo {volume} {55}},\ \bibinfo {pages} {271} (\bibinfo {year}
  {2005})},\ \Eprint {http://arxiv.org/abs/nucl-ex/0502005}
  {arXiv:nucl-ex/0502005} \BibitemShut {NoStop}%
\bibitem [{\citenamefont {Baltz}(2008)}]{Baltz:2007kq}%
  \BibitemOpen
  \bibfield  {author} {\bibinfo {author} {\bibfnamefont {A.~J.}\ \bibnamefont
  {Baltz}},\ }\href {\doibase 10.1016/j.physrep.2007.12.001} {\bibfield
  {journal} {\bibinfo  {journal} {Phys. Rept.}\ }\textbf {\bibinfo {volume}
  {458}},\ \bibinfo {pages} {1} (\bibinfo {year} {2008})},\ \Eprint
  {http://arxiv.org/abs/0706.3356} {arXiv:0706.3356 [nucl-ex]} \BibitemShut
  {NoStop}%
\bibitem [{\citenamefont {Contreras}\ and\ \citenamefont
  {Tapia~Takaki}(2015)}]{Contreras:2015dqa}%
  \BibitemOpen
  \bibfield  {author} {\bibinfo {author} {\bibfnamefont {J.~G.}\ \bibnamefont
  {Contreras}}\ and\ \bibinfo {author} {\bibfnamefont {J.~D.}\ \bibnamefont
  {Tapia~Takaki}},\ }\href {\doibase 10.1142/S0217751X15420129} {\bibfield
  {journal} {\bibinfo  {journal} {Int. J. Mod. Phys. A}\ }\textbf {\bibinfo
  {volume} {30}},\ \bibinfo {pages} {1542012} (\bibinfo {year}
  {2015})}\BibitemShut {NoStop}%
\bibitem [{\citenamefont {Accardi}\ \emph {et~al.}(2016)\citenamefont {Accardi}
  \emph {et~al.}}]{Accardi:2012qut}%
  \BibitemOpen
  \bibfield  {author} {\bibinfo {author} {\bibfnamefont {A.}~\bibnamefont
  {Accardi}} \emph {et~al.},\ }\href {\doibase 10.1140/epja/i2016-16268-9}
  {\bibfield  {journal} {\bibinfo  {journal} {Eur. Phys. J. A}\ }\textbf
  {\bibinfo {volume} {52}},\ \bibinfo {pages} {268} (\bibinfo {year} {2016})},\
  \Eprint {http://arxiv.org/abs/1212.1701} {arXiv:1212.1701 [nucl-ex]}
  \BibitemShut {NoStop}%
\bibitem [{\citenamefont {Abdul~Khalek}\ \emph {et~al.}(2021)\citenamefont
  {Abdul~Khalek} \emph {et~al.}}]{AbdulKhalek:2021gbh}%
  \BibitemOpen
  \bibfield  {author} {\bibinfo {author} {\bibfnamefont {R.}~\bibnamefont
  {Abdul~Khalek}} \emph {et~al.},\ }\href@noop {} {\  (\bibinfo {year}
  {2021})},\ \Eprint {http://arxiv.org/abs/2103.05419} {arXiv:2103.05419
  [physics.ins-det]} \BibitemShut {NoStop}%
\bibitem [{\citenamefont {Adamczyk}\ \emph {et~al.}(2017)\citenamefont
  {Adamczyk} \emph {et~al.}}]{STAR:2017enh}%
  \BibitemOpen
  \bibfield  {author} {\bibinfo {author} {\bibfnamefont {L.}~\bibnamefont
  {Adamczyk}} \emph {et~al.} (\bibinfo {collaboration} {STAR}),\ }\href
  {\doibase 10.1103/PhysRevC.96.054904} {\bibfield  {journal} {\bibinfo
  {journal} {Phys. Rev. C}\ }\textbf {\bibinfo {volume} {96}},\ \bibinfo
  {pages} {054904} (\bibinfo {year} {2017})},\ \Eprint
  {http://arxiv.org/abs/1702.07705} {arXiv:1702.07705 [nucl-ex]} \BibitemShut
  {NoStop}%
\bibitem [{\citenamefont {Klein}\ and\ \citenamefont
  {M\"antysaari}(2019)}]{Klein:2019qfb}%
  \BibitemOpen
  \bibfield  {author} {\bibinfo {author} {\bibfnamefont {S.~R.}\ \bibnamefont
  {Klein}}\ and\ \bibinfo {author} {\bibfnamefont {H.}~\bibnamefont
  {M\"antysaari}},\ }\href {\doibase 10.1038/s42254-019-0107-6} {\bibfield
  {journal} {\bibinfo  {journal} {Nature Rev. Phys.}\ }\textbf {\bibinfo
  {volume} {1}},\ \bibinfo {pages} {662} (\bibinfo {year} {2019})},\ \Eprint
  {http://arxiv.org/abs/1910.10858} {arXiv:1910.10858 [hep-ex]} \BibitemShut
  {NoStop}%
\bibitem [{\citenamefont {Good}\ and\ \citenamefont
  {Walker}(1960)}]{Good:1960ba}%
  \BibitemOpen
  \bibfield  {author} {\bibinfo {author} {\bibfnamefont {M.~L.}\ \bibnamefont
  {Good}}\ and\ \bibinfo {author} {\bibfnamefont {W.~D.}\ \bibnamefont
  {Walker}},\ }\href {\doibase 10.1103/PhysRev.120.1857} {\bibfield  {journal}
  {\bibinfo  {journal} {Phys. Rev.}\ }\textbf {\bibinfo {volume} {120}},\
  \bibinfo {pages} {1857} (\bibinfo {year} {1960})}\BibitemShut {NoStop}%
\bibitem [{\citenamefont {Miettinen}\ and\ \citenamefont
  {Pumplin}(1978)}]{Miettinen:1978jb}%
  \BibitemOpen
  \bibfield  {author} {\bibinfo {author} {\bibfnamefont {H.~I.}\ \bibnamefont
  {Miettinen}}\ and\ \bibinfo {author} {\bibfnamefont {J.}~\bibnamefont
  {Pumplin}},\ }\href {\doibase 10.1103/PhysRevD.18.1696} {\bibfield  {journal}
  {\bibinfo  {journal} {Phys. Rev. D}\ }\textbf {\bibinfo {volume} {18}},\
  \bibinfo {pages} {1696} (\bibinfo {year} {1978})}\BibitemShut {NoStop}%
\bibitem [{\citenamefont {Frankfurt}\ \emph {et~al.}(2022)\citenamefont
  {Frankfurt}, \citenamefont {Guzey}, \citenamefont {Stasto},\ and\
  \citenamefont {Strikman}}]{Frankfurt:2022jns}%
  \BibitemOpen
  \bibfield  {author} {\bibinfo {author} {\bibfnamefont {L.}~\bibnamefont
  {Frankfurt}}, \bibinfo {author} {\bibfnamefont {V.}~\bibnamefont {Guzey}},
  \bibinfo {author} {\bibfnamefont {A.}~\bibnamefont {Stasto}}, \ and\ \bibinfo
  {author} {\bibfnamefont {M.}~\bibnamefont {Strikman}},\ }\href@noop {} {\
  (\bibinfo {year} {2022})},\ \Eprint {http://arxiv.org/abs/2203.12289}
  {arXiv:2203.12289 [hep-ph]} \BibitemShut {NoStop}%
\bibitem [{\citenamefont {Adler}\ \emph {et~al.}(2002)\citenamefont {Adler}
  \emph {et~al.}}]{STAR:2002caw}%
  \BibitemOpen
  \bibfield  {author} {\bibinfo {author} {\bibfnamefont {C.}~\bibnamefont
  {Adler}} \emph {et~al.} (\bibinfo {collaboration} {STAR}),\ }\href {\doibase
  10.1103/PhysRevLett.89.272302} {\bibfield  {journal} {\bibinfo  {journal}
  {Phys. Rev. Lett.}\ }\textbf {\bibinfo {volume} {89}},\ \bibinfo {pages}
  {272302} (\bibinfo {year} {2002})},\ \Eprint
  {http://arxiv.org/abs/nucl-ex/0206004} {arXiv:nucl-ex/0206004} \BibitemShut
  {NoStop}%
\bibitem [{\citenamefont {Abelev}\ \emph {et~al.}(2008)\citenamefont {Abelev}
  \emph {et~al.}}]{STAR:2007elq}%
  \BibitemOpen
  \bibfield  {author} {\bibinfo {author} {\bibfnamefont {B.~I.}\ \bibnamefont
  {Abelev}} \emph {et~al.} (\bibinfo {collaboration} {STAR}),\ }\href {\doibase
  10.1103/PhysRevC.77.034910} {\bibfield  {journal} {\bibinfo  {journal} {Phys.
  Rev. C}\ }\textbf {\bibinfo {volume} {77}},\ \bibinfo {pages} {034910}
  (\bibinfo {year} {2008})},\ \Eprint {http://arxiv.org/abs/0712.3320}
  {arXiv:0712.3320 [nucl-ex]} \BibitemShut {NoStop}%
\bibitem [{\citenamefont {Abelev}\ \emph {et~al.}(2010)\citenamefont {Abelev}
  \emph {et~al.}}]{STAR:2009giy}%
  \BibitemOpen
  \bibfield  {author} {\bibinfo {author} {\bibfnamefont {B.~I.}\ \bibnamefont
  {Abelev}} \emph {et~al.} (\bibinfo {collaboration} {STAR}),\ }\href {\doibase
  10.1103/PhysRevC.81.044901} {\bibfield  {journal} {\bibinfo  {journal} {Phys.
  Rev. C}\ }\textbf {\bibinfo {volume} {81}},\ \bibinfo {pages} {044901}
  (\bibinfo {year} {2010})},\ \Eprint {http://arxiv.org/abs/0912.0604}
  {arXiv:0912.0604 [nucl-ex]} \BibitemShut {NoStop}%
\bibitem [{\citenamefont {Acharya}\ \emph {et~al.}(2020)\citenamefont {Acharya}
  \emph {et~al.}}]{ALICE:2020ugp}%
  \BibitemOpen
  \bibfield  {author} {\bibinfo {author} {\bibfnamefont {S.}~\bibnamefont
  {Acharya}} \emph {et~al.} (\bibinfo {collaboration} {ALICE}),\ }\href
  {\doibase 10.1007/JHEP06(2020)035} {\bibfield  {journal} {\bibinfo  {journal}
  {JHEP}\ }\textbf {\bibinfo {volume} {06}},\ \bibinfo {pages} {035} (\bibinfo
  {year} {2020})},\ \Eprint {http://arxiv.org/abs/2002.10897} {arXiv:2002.10897
  [nucl-ex]} \BibitemShut {NoStop}%
\bibitem [{\citenamefont {Adam}\ \emph {et~al.}(2019)\citenamefont {Adam} \emph
  {et~al.}}]{STAR:2019yox}%
  \BibitemOpen
  \bibfield  {author} {\bibinfo {author} {\bibfnamefont {J.}~\bibnamefont
  {Adam}} \emph {et~al.} (\bibinfo {collaboration} {STAR}),\ }\href {\doibase
  10.1103/PhysRevLett.123.132302} {\bibfield  {journal} {\bibinfo  {journal}
  {Phys. Rev. Lett.}\ }\textbf {\bibinfo {volume} {123}},\ \bibinfo {pages}
  {132302} (\bibinfo {year} {2019})},\ \Eprint
  {http://arxiv.org/abs/1904.11658} {arXiv:1904.11658 [hep-ex]} \BibitemShut
  {NoStop}%
\bibitem [{\citenamefont {{ALICE Collaboration}}(2022)}]{ALICE:2022zso}%
  \BibitemOpen
  \bibfield  {author} {\bibinfo {author} {\bibnamefont {{ALICE Collaboration}}}
  (\bibinfo {collaboration} {ALICE}),\ }\href@noop {} {\  (\bibinfo {year}
  {2022})},\ \Eprint {http://arxiv.org/abs/2204.10684} {arXiv:2204.10684
  [nucl-ex]} \BibitemShut {NoStop}%
\bibitem [{\citenamefont {Aaij}\ \emph {et~al.}(2022)\citenamefont {Aaij} \emph
  {et~al.}}]{LHCb:2021hoq}%
  \BibitemOpen
  \bibfield  {author} {\bibinfo {author} {\bibfnamefont {R.}~\bibnamefont
  {Aaij}} \emph {et~al.} (\bibinfo {collaboration} {LHCb}),\ }\href {\doibase
  10.1103/PhysRevC.105.L032201} {\bibfield  {journal} {\bibinfo  {journal}
  {Phys. Rev. C}\ }\textbf {\bibinfo {volume} {105}},\ \bibinfo {pages}
  {L032201} (\bibinfo {year} {2022})},\ \Eprint
  {http://arxiv.org/abs/2108.02681} {arXiv:2108.02681 [hep-ex]} \BibitemShut
  {NoStop}%
\bibitem [{\citenamefont {Miettinen}\ and\ \citenamefont
  {Pumplin}(1979)}]{Miettinen:1978xv}%
  \BibitemOpen
  \bibfield  {author} {\bibinfo {author} {\bibfnamefont {H.~I.}\ \bibnamefont
  {Miettinen}}\ and\ \bibinfo {author} {\bibfnamefont {J.}~\bibnamefont
  {Pumplin}},\ }\href {\doibase 10.1103/PhysRevLett.42.204} {\bibfield
  {journal} {\bibinfo  {journal} {Phys. Rev. Lett.}\ }\textbf {\bibinfo
  {volume} {42}},\ \bibinfo {pages} {204} (\bibinfo {year} {1979})}\BibitemShut
  {NoStop}%
\bibitem [{\citenamefont {Acharya}\ \emph {et~al.}(2021)\citenamefont {Acharya}
  \emph {et~al.}}]{ALICE:2021gpt}%
  \BibitemOpen
  \bibfield  {author} {\bibinfo {author} {\bibfnamefont {S.}~\bibnamefont
  {Acharya}} \emph {et~al.} (\bibinfo {collaboration} {ALICE}),\ }\href
  {\doibase 10.1140/epjc/s10052-021-09437-6} {\bibfield  {journal} {\bibinfo
  {journal} {Eur. Phys. J. C}\ }\textbf {\bibinfo {volume} {81}},\ \bibinfo
  {pages} {712} (\bibinfo {year} {2021})},\ \Eprint
  {http://arxiv.org/abs/2101.04577} {arXiv:2101.04577 [nucl-ex]} \BibitemShut
  {NoStop}%
\bibitem [{\citenamefont {Gribov}(1969)}]{Gribov:1968jf}%
  \BibitemOpen
  \bibfield  {author} {\bibinfo {author} {\bibfnamefont {V.~N.}\ \bibnamefont
  {Gribov}},\ }\href@noop {} {\bibfield  {journal} {\bibinfo  {journal} {Sov.
  Phys. JETP}\ }\textbf {\bibinfo {volume} {29}},\ \bibinfo {pages} {483}
  (\bibinfo {year} {1969})}\BibitemShut {NoStop}%
\bibitem [{\citenamefont {Klein}\ and\ \citenamefont
  {Nystrand}(1999)}]{Klein:1999qj}%
  \BibitemOpen
  \bibfield  {author} {\bibinfo {author} {\bibfnamefont {S.}~\bibnamefont
  {Klein}}\ and\ \bibinfo {author} {\bibfnamefont {J.}~\bibnamefont
  {Nystrand}},\ }\href {\doibase 10.1103/PhysRevC.60.014903} {\bibfield
  {journal} {\bibinfo  {journal} {Phys. Rev. C}\ }\textbf {\bibinfo {volume}
  {60}},\ \bibinfo {pages} {014903} (\bibinfo {year} {1999})},\ \Eprint
  {http://arxiv.org/abs/hep-ph/9902259} {arXiv:hep-ph/9902259} \BibitemShut
  {NoStop}%
\bibitem [{\citenamefont {Frankfurt}\ \emph {et~al.}(2003)\citenamefont
  {Frankfurt}, \citenamefont {Strikman},\ and\ \citenamefont
  {Zhalov}}]{Frankfurt:2002sv}%
  \BibitemOpen
  \bibfield  {author} {\bibinfo {author} {\bibfnamefont {L.}~\bibnamefont
  {Frankfurt}}, \bibinfo {author} {\bibfnamefont {M.}~\bibnamefont {Strikman}},
  \ and\ \bibinfo {author} {\bibfnamefont {M.}~\bibnamefont {Zhalov}},\ }\href
  {\doibase 10.1103/PhysRevC.67.034901} {\bibfield  {journal} {\bibinfo
  {journal} {Phys. Rev. C}\ }\textbf {\bibinfo {volume} {67}},\ \bibinfo
  {pages} {034901} (\bibinfo {year} {2003})},\ \Eprint
  {http://arxiv.org/abs/hep-ph/0210303} {arXiv:hep-ph/0210303} \BibitemShut
  {NoStop}%
\bibitem [{\citenamefont {Frankfurt}\ \emph {et~al.}(2016)\citenamefont
  {Frankfurt}, \citenamefont {Guzey}, \citenamefont {Strikman},\ and\
  \citenamefont {Zhalov}}]{Frankfurt:2015cwa}%
  \BibitemOpen
  \bibfield  {author} {\bibinfo {author} {\bibfnamefont {L.}~\bibnamefont
  {Frankfurt}}, \bibinfo {author} {\bibfnamefont {V.}~\bibnamefont {Guzey}},
  \bibinfo {author} {\bibfnamefont {M.}~\bibnamefont {Strikman}}, \ and\
  \bibinfo {author} {\bibfnamefont {M.}~\bibnamefont {Zhalov}},\ }\href
  {\doibase 10.1016/j.physletb.2015.11.012} {\bibfield  {journal} {\bibinfo
  {journal} {Phys. Lett. B}\ }\textbf {\bibinfo {volume} {752}},\ \bibinfo
  {pages} {51} (\bibinfo {year} {2016})},\ \Eprint
  {http://arxiv.org/abs/1506.07150} {arXiv:1506.07150 [hep-ph]} \BibitemShut
  {NoStop}%
\bibitem [{\citenamefont {Guzey}\ \emph {et~al.}(2013)\citenamefont {Guzey},
  \citenamefont {Kryshen}, \citenamefont {Strikman},\ and\ \citenamefont
  {Zhalov}}]{Guzey:2013xba}%
  \BibitemOpen
  \bibfield  {author} {\bibinfo {author} {\bibfnamefont {V.}~\bibnamefont
  {Guzey}}, \bibinfo {author} {\bibfnamefont {E.}~\bibnamefont {Kryshen}},
  \bibinfo {author} {\bibfnamefont {M.}~\bibnamefont {Strikman}}, \ and\
  \bibinfo {author} {\bibfnamefont {M.}~\bibnamefont {Zhalov}},\ }\href
  {\doibase 10.1016/j.physletb.2013.08.043} {\bibfield  {journal} {\bibinfo
  {journal} {Phys. Lett. B}\ }\textbf {\bibinfo {volume} {726}},\ \bibinfo
  {pages} {290} (\bibinfo {year} {2013})},\ \Eprint
  {http://arxiv.org/abs/1305.1724} {arXiv:1305.1724 [hep-ph]} \BibitemShut
  {NoStop}%
\bibitem [{\citenamefont {Zha}\ \emph {et~al.}(2018)\citenamefont {Zha},
  \citenamefont {Klein}, \citenamefont {Ma}, \citenamefont {Ruan},
  \citenamefont {Todoroki}, \citenamefont {Tang}, \citenamefont {Xu},
  \citenamefont {Yang}, \citenamefont {Yang},\ and\ \citenamefont
  {Yang}}]{Zha:2017jch}%
  \BibitemOpen
  \bibfield  {author} {\bibinfo {author} {\bibfnamefont {W.}~\bibnamefont
  {Zha}}, \bibinfo {author} {\bibfnamefont {S.~R.}\ \bibnamefont {Klein}},
  \bibinfo {author} {\bibfnamefont {R.}~\bibnamefont {Ma}}, \bibinfo {author}
  {\bibfnamefont {L.}~\bibnamefont {Ruan}}, \bibinfo {author} {\bibfnamefont
  {T.}~\bibnamefont {Todoroki}}, \bibinfo {author} {\bibfnamefont
  {Z.}~\bibnamefont {Tang}}, \bibinfo {author} {\bibfnamefont {Z.}~\bibnamefont
  {Xu}}, \bibinfo {author} {\bibfnamefont {C.}~\bibnamefont {Yang}}, \bibinfo
  {author} {\bibfnamefont {Q.}~\bibnamefont {Yang}}, \ and\ \bibinfo {author}
  {\bibfnamefont {S.}~\bibnamefont {Yang}},\ }\href {\doibase
  10.1103/PhysRevC.97.044910} {\bibfield  {journal} {\bibinfo  {journal} {Phys.
  Rev. C}\ }\textbf {\bibinfo {volume} {97}},\ \bibinfo {pages} {044910}
  (\bibinfo {year} {2018})},\ \Eprint {http://arxiv.org/abs/1705.01460}
  {arXiv:1705.01460 [nucl-th]} \BibitemShut {NoStop}%
\bibitem [{\citenamefont {K\l{}usek-Gawenda}\ and\ \citenamefont
  {Szczurek}(2016)}]{Klusek-Gawenda:2015hja}%
  \BibitemOpen
  \bibfield  {author} {\bibinfo {author} {\bibfnamefont {M.}~\bibnamefont
  {K\l{}usek-Gawenda}}\ and\ \bibinfo {author} {\bibfnamefont {A.}~\bibnamefont
  {Szczurek}},\ }\href {\doibase 10.1103/PhysRevC.93.044912} {\bibfield
  {journal} {\bibinfo  {journal} {Phys. Rev. C}\ }\textbf {\bibinfo {volume}
  {93}},\ \bibinfo {pages} {044912} (\bibinfo {year} {2016})},\ \Eprint
  {http://arxiv.org/abs/1509.03173} {arXiv:1509.03173 [nucl-th]} \BibitemShut
  {NoStop}%
\bibitem [{\citenamefont {Klein}(2022)}]{Klein:2021mgd}%
  \BibitemOpen
  \bibfield  {author} {\bibinfo {author} {\bibfnamefont {S.~R.}\ \bibnamefont
  {Klein}} (\bibinfo {collaboration} {STAR}),\ }\href {\doibase
  10.21468/SciPostPhysProc.8.128} {\bibfield  {journal} {\bibinfo  {journal}
  {SciPost Phys. Proc.}\ }\textbf {\bibinfo {volume} {8}},\ \bibinfo {pages}
  {128} (\bibinfo {year} {2022})},\ \Eprint {http://arxiv.org/abs/2107.10447}
  {arXiv:2107.10447 [nucl-ex]} \BibitemShut {NoStop}%
\bibitem [{\citenamefont {M\"antysaari}\ \emph {et~al.}(2022)\citenamefont
  {M\"antysaari}, \citenamefont {Salazar},\ and\ \citenamefont
  {Schenke}}]{Mantysaari:2022sux}%
  \BibitemOpen
  \bibfield  {author} {\bibinfo {author} {\bibfnamefont {H.}~\bibnamefont
  {M\"antysaari}}, \bibinfo {author} {\bibfnamefont {F.}~\bibnamefont
  {Salazar}}, \ and\ \bibinfo {author} {\bibfnamefont {B.}~\bibnamefont
  {Schenke}},\ }\href {\doibase 10.1103/PhysRevD.106.074019} {\bibfield
  {journal} {\bibinfo  {journal} {Phys. Rev. D}\ }\textbf {\bibinfo {volume}
  {106}},\ \bibinfo {pages} {074019} (\bibinfo {year} {2022})},\ \Eprint
  {http://arxiv.org/abs/2207.03712} {arXiv:2207.03712 [hep-ph]} \BibitemShut
  {NoStop}%
\bibitem [{\citenamefont {IAEA}(2022)}]{levels}%
  \BibitemOpen
  \bibfield  {author} {\bibinfo {author} {\bibnamefont {IAEA}},\ }\href@noop {}
  {\enquote {\bibinfo {title} {Live chart of nuclides},}\ }\bibinfo
  {howpublished} {\url{
  https://nds.iaea.org/relnsd/vcharthtml/VChartHTML.html}} (\bibinfo {year}
  {Accessed Nov., 2022})\BibitemShut {NoStop}%
\end{thebibliography}%
\end{document}